\documentclass[pra,10pt,british,aps,reprint]{revtex4-1}

\usepackage[T1]{fontenc}
\usepackage[latin9]{inputenc}
\usepackage[letterpaper]{geometry}
\usepackage{amsmath}
\usepackage{graphicx}
\usepackage{esint}



\usepackage{babel}
\begin{document}

\title{Coherent Control of Photoelectron Wavepacket Angular Interferograms}

\author{P. Hockett}

\email{paul.hockett@nrc.ca}

\affiliation{National Research Council of Canada, 100 Sussex Drive, Ottawa,K1A
0R6, Canada}

\author{M. Wollenhaupt}

\affiliation{Institut für Physik, Carl von Ossietzky Universität Oldenburg, Carl-von-Ossietzky-Straße
9-11, 26129 Oldenburg, Germany}

\author{T. Baumert}

\affiliation{Institut für Physik, Universität Kassel, Heinrich-Plett-Str. 40,
34132 Kassel, Germany}

\begin{abstract}
Coherent control over photoelectron wavepackets, via the use of polarization-shaped laser pulses, can be understood as a time and polarization-multiplexed process. In this work, we investigate this multiplexing via computation of the observable photoelectron angular interferograms resulting from multi-photon atomic ionization with polarization-shaped laser pulses. We consider the polarization sensitivity of both the instantaneous and cumulative continuum wavefunction; the nature of the coherent control over the resultant photoelectron interferogram is thus explored in detail. Based on this understanding, the use of coherent control with polarization-shaped pulses as a methodology for a highly multiplexed coherent quantum metrology is also investigated, and defined in terms of the information content of the observable.
\end{abstract}

\maketitle

\section{Introduction}

In the area of coherent control the coherence properties of light
together with quantum-mechanical matter interferences are used to
steer a quantum system to a desired target or dynamical behavior.
While original ideas were developed in the physical-chemistry community,
the field of quantum control has grown well beyond its traditional
boundaries, and a tremendous cross-fertilization to neighboring quantum
technologies in terms of both experimental techniques and theoretical
developments has occurred \cite{Shapiro2011}. The increasing availability
of laser sources operating on the time scale of molecular dynamics,
i.e. the femtosecond regime, and the increasing capabilities of shaping
light in terms of amplitude, phase and polarization - also on the
time scale of molecular dynamics - brought the temporal aspect of
this field to the fore (see for example ref. \cite{Wollenhaupt2011}
and references therein). 

A thoroughly investigated control regime is the interference between
N and M photons \cite{Shapiro2011}. Taking into account selection
rules, there are in general terms two types of scenario. In cases
where N and M are both either odd or even, the integral as well as
the differential cross section can be controlled as a function of
phase between the N photon and M photon field. In cases where N and
M have different parity, only the differential cross section, i.e.
a scattering process into different angles, can be controlled. In
the following we focus on the N and N case. Using two-photon transitions,
the transition from 6s$\rightarrow$7d in cesium was studied in two
regimes, spanning from optical interferences to quantum interferences,
by Girard et al. \cite{Blanchet1997}. Using pulse shaping technologies,
this scheme has been exploited many times and in the perturbative
interaction regime the physical process can be related to (higher
order) spectral interferences \cite{Meshulach1998,Prakelt2004,Walowicz2002,Ruge2013}.
In the non-perturbative interaction regime for one photon transitions
control via selective population of dressed states was demonstrated,
originally on atoms \cite{Wollenhaupt2003b} and, later, used for
control of the coupled nuclear and electronic dynamics in molecules
\cite{Bayer2013}. For the one photon perturbative case the transition
from optical to quantum interferences was studied on the 4s$\rightarrow$4p
transition in potassium by Girard's group \cite{Bouchene1998} and
the same group studied later the coherent buildup of population in
Rb (5s$\rightarrow$5p) during chirped excitation \cite{Zamith2001}.
It turned out that these coherent transients can be used for quantum
state measurements \cite{Monmayrant2006}. 

These investigations were extended to bound-free transitions and the
interference of free electron wave packets on threshold electrons
as well as on the first ATI channel \cite{Wollenhaupt2002} was studied
on potassium atoms, a time domain analogue to Young\textquoteright{}s
double slit experiment. Later these interference investigations were
extended to three photon ionization processes with polarization shaped
laser pulses \cite{Wollenhaupt2009a,Wollenhaupt2013}, tomographic
reconstruction techniques were developed to measure the three dimensional
photoelectron momentum distributions, and applied to measurements
of a 1+2 photon resonantly-enhanced ionization process in potassium
atoms \cite{Wollenhaupt2009} and to ionization of chiral molecules
\cite{Lux2015}. 

It was then realized, that these multiplexed data can be used to determine
the radial phase shifts of the ionization matrix elements where dynamics
in the matter system and in the ionizing pathways were exploited \cite{Hockett2014}.
In this contribution we take the topic of complete ionization studies
as an example that coherent control in the time domain with polarization
shaped pulses represents a powerful tool to metrology. We study the
transient build up of the final continuum interference and argue that
due to the coherent nature the information on ionization matrix elements
can be extracted in a much shorter measurement time as compared to
serial ionization schemes with time-independent polarization states.

\section{Coherent control over photoelectron interferograms}

\begin{figure}
\includegraphics[scale=0.85]{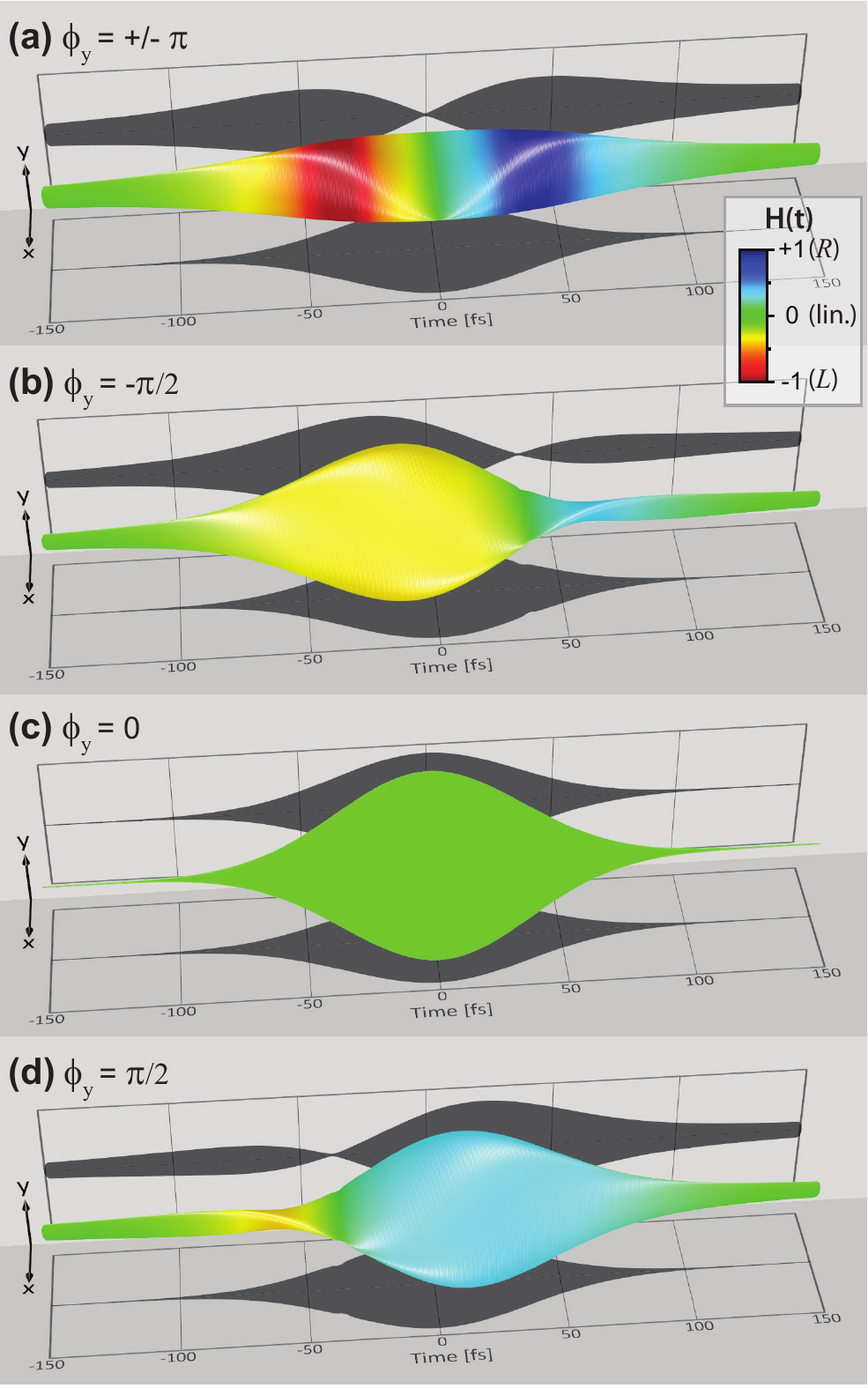}
\caption{
Polarization-shaped pulses $E(x,y,t)$, resulting from a phase step
applied to the spectral phase, as defined by eqn. \ref{eq:phi_y}.
The helicity, $H(t)$ (eqn. \ref{eq:helicity}), is shown by the colour-map.
The photoelectron interferograms resulting from these pulses are shown
in figure \ref{fig:Photoelectron-interferograms-4up}.\label{fig:Polarization-shaped-pulses}
}
\end{figure}

Many types of interference can be manifested in continuum photoelectron
wavepackets. The simplest case, time-independent photoionization at
a single energy, results in an outgoing photoelectron wave with an
angular structure defined by the interferences between different angular
momentum components (\emph{partial waves}); this phenomenon, and control
over the resulting interferograms, has long been investigated in the
energy-domain \cite{Cooper1968,Yin1992,Yin1995}. In the time-domain,
the possibility of preparing and controlling single or multiple photoelectron
wavepackets, allows for the effects of coherence to appear in both
the angular and energy structure of the observable photoelectron flux,
and thus further possibilities for coherent control methodologies
\cite{Wollenhaupt2002,Wollenhaupt2003b,Wollenhaupt2009a}. Regardless
of the exact scheme employed, the light-matter interaction involves
ionization of a target system, with some form of control over this
process via the laser field applied. In the most basic case, control
is achieved via the ionization dynamics, in essence by affecting the
magnitudes and/or phases of the accessible ionization pathways; in
more complex cases the interaction may additionally involve dynamics
in the matter system (which may be laser-driven or inherent to the
system), and continuum dynamics which occur post-ionization.

In this work, we discuss the specific case of coherent control over
the angular structure of a photoelectron wavepacket, by making use
of ultrafast polarization-shaped pulses interacting with, and ultimately
ionizing, potassium atoms. This particular scheme incorporates both
intra-pulse bound-state dynamics, driven by the moderately intense
laser field, and net 3-photon ionization. The scheme is outlined in
some detail in refs. \cite{Wollenhaupt2002,Wollenhaupt2009a}, and
more recently in refs. \cite{Hockett2014,Hockett2015a}, wherein we
focussed primarily on determining the details of the interaction,
in particular the determination of the ionization matrix elements
from experimental measurements. We focus herein on the coherent control
possible with polarization-shaped pulses, making use of the previously
determined matrix elements in our calculations, and investigate the
time-dependent aspects of the resulting interferograms and their further
use for metrology.

We begin with an overview of the details pertinent to the consideration
of polarization-shaped pulses. The laser pulse is described generally
by the electric field $E(x,y,t)$, i.e. in a Cartesian basis and propagating
in the $z$-direction. This field can be defined in terms of its spectrum
$\tilde{E}(\Omega)$ and associated spectral phase $\phi$, which
we allow to be independent for the two polarization components:

\begin{equation}
\left(\begin{array}{c}
E_{x}(t)\\
E_{y}(t)
\end{array}\right)=\mathcal{F}^{-1}\left\{ \tilde{E}(\Omega)\left(\begin{array}{c}
e^{i\phi_{x}(\Omega)}\\
e^{i\phi_{y}(\Omega)}
\end{array}\right)\right\} \label{eq:E-x-y}
\end{equation}

In this form, shaping of the polarization of the field is defined
by the spectral phases $\phi_{x}(\Omega)$ and $\phi_{y}(\Omega)$.
A field with $\phi_{x}=\phi_{y}$ over the full spectrum will be linearly
polarized; a field with a frequency-independent phase shift ($\phi_{x}\neq\phi_{y}$)
will be in a ``pure'' elliptical or circular state, with no time-dependence
of the polarization; a field with a frequency-\emph{dependent} phase
shift will produce a fully polarization-shaped pulse, with a complex
temporal-dependence of the polarization and spectral content of the
pulse \cite{Brixner2001}. For the shaped pulses explored in this
manuscript, the spectral phase is set to a single value for the red-half
of the pulse and $y$-component only, and is zero elsewhere, i.e.
there is a phase step defined as:

\begin{eqnarray}
\phi_{y}(\Omega) & = & \begin{cases}
\phi_{y}, & \Omega<\Omega_{0}\\
0, & \Omega\geq\Omega_{0}
\end{cases}\label{eq:phi_y}\\
\phi_{x}(\Omega) & = & 0
\end{eqnarray}
where $\Omega_{0}$ is the central frequency of the pulse. Within
this definition, the full spectral phase function $\phi_{y}(\Omega)$
is parametrized by the single value $\phi_{y}$, and this value is
used to label the shaped pulses in this work. (For more general discussion
of pulse-shapes arising from spectral phase steps, the reader is referred
to ref. \cite{Wollenhaupt2010}.)

For the case of a polarization-shaped pulse, it is convenient to express
the pulse in terms of left and right circularly polarized components,
$E_{L}(t)$ and $E_{R}(t)$. For atomic ionization this is also a
good choice of basis, since the left and right circularly polarized
components will couple to different $m$-states, specifically to $-m$
and $+m$ states respectively \cite{Hockett2014}.

\begin{equation}
\left(\begin{array}{c}
E_{L}(t)\\
E_{R}(t)
\end{array}\right)=\frac{1}{\sqrt{2}}\left(\begin{array}{c}
E_{x}(t)-iE_{y}(t)\\
E_{x}(t)+iE_{y}(t)
\end{array}\right)\label{eq:E-spherical}
\end{equation}

In this basis we can also express the helicity of the pulse, which
we define as the normalized difference between the $E_{L}(t)$ and
$E_{R}(t)$ components:

\begin{equation}
H(t)=\frac{|E_{R}(t)|-|E_{L}(t)|}{|E_{R}(t)|+|E_{L}(t)|}\label{eq:helicity}
\end{equation}
Hence, $H(t)=1$ for a field which is right circularly polarized
at time $t$, $H(t)=-1$ for a left circularly polarized field, and
$H(t)=0$ for a linearly polarized field. All other values define
varying degrees of ellipticity, with an overall right or left handedness
for positive or negative helicities respectively. The behaviour of
$H(t)$ is thus equivalent to the (normalized) $S_{3}(t)$ Stokes
parameter, which also parametrizes the degree of $R$ or $L$ polarization
and takes values on the interval $+1\geq S_{3}\geq-1$ \cite{Guenther1990},
but is defined directly from the $L/R$ electric field basis.

Examples of the pulse shapes arising from spectral phase steps, as
defined by eqn. \ref{eq:phi_y}, are given in figure \ref{fig:Polarization-shaped-pulses}.
In all cases the pulse is initially defined to be Gaussian, with bandwidth
matching the transform-limited pulse duration of $\tau_{FWHM}=60$~fs,
and is linearly polarized. The application of a phase-step, with value
$\phi_{y}$, results in a temporally varying polarization state as
discussed above, with the pulse passing through various degrees of
ellipticity as a function of the magnitudes and phases of the $E_{L}(t)$
and $E_{R}(t)$ components. The colour-map in figure \ref{fig:Polarization-shaped-pulses}
shows this time-dependence in terms of the helicity $H(t)$ (eqn.
\ref{eq:helicity}), which is most directly related to the contributing
ionization pathways

For the net 3-photon ionization of potassium, which we use here as
our model system since the ionization dynamics have been determined
\cite{Hockett2014}, the overall process to a final state $f$ can
be written as:

\begin{equation}
d_{f}(\mathbf{k},t)=\sum_{i,v}d_{i\rightarrow v}(\mathbf{k},t)d_{v\rightarrow f}(\mathbf{k},t)\chi_{i}(t)
\end{equation}
where $d_{f}(\mathbf{k},t)$ is the effective 2-photon ionization
dipole moment, defined as a product of two 1-photon terms (from an
initial state $i$ to a final state $f$ via a virtual state $v$)
and the ionizable bound-state population $\chi_{i}(t)$. In this case,
the ionizable bound-state population comprises the excited $4p$ states,
which are strongly-coupled to the ground $4s$ states by the near-resonant
laser field. Thus, the bound-state dynamics are driven by the laser
field, and play a significant role in the accessible final states. 

The final observable photoelectron interferogram is
given by the coherent square over all final continuum states, and can be written as:

\begin{widetext}
\begin{eqnarray}
I(\theta,\phi;\, k) & = & \iint dk\, dt\sum_{f,f'}d_{f}(\mathbf{k},t)d_{f'}^{*}(\mathbf{k},t)\\
 & = & \iint dk\, dt\sum_{\begin{array}{c}
l,m\\
l',m'
\end{array}}d_{l,m}(k,t)Y_{l,m}(\theta,\phi)d_{l',m'}^{*}(k,t)Y_{l',m'}^{*}(\theta,\phi)\\
 & \equiv & \iint dk\, dt\sum_{\begin{array}{c}
l,m\\
l',m'
\end{array}}\psi_{l,m}(\mathbf{k},t)\psi_{l',m'}^{*}(\mathbf{k},t)
\end{eqnarray} \end{widetext}
Here the final continuum states are expressed in terms of angular
momentum states $(l,\, m)$, with angular dependence given
by spherical harmonics $Y_{l,m}(\theta,\phi)$. Equivalently, the
expression can be given in terms of the components of the continuum
photoelectron wavefunction $\Psi(\mathbf{k},t)=\sum_{l,m}\psi_{l,m}(\mathbf{k},t)$,
which most clearly define the origin of the observable photoelectron
interferogram. The final energy and angle-resolved observable is the
coherent square of this continuum wavefunction, integrated over time
$t$ and photoelectron energy $k$, for
a small energy range $dk$ over which we assume the ionization dipole
moments $d_{l,m}(k,t)$ are constant. For a more detailed discussion
and derivation the reader is referred to ref. \cite{Hockett2015a}.

In terms of control, the details of the laser field affect both the
bound-state populations $\chi_{i}(t)$, and their coupling to the
final continuum states $\psi_{l,m}(\mathbf{k},t)$.
This is particularly clear in the $E_{L/R}$ basis, since these components
drive the final population towards $m<0$ and $m>0$ states respectively
(see refs. \cite{Hockett2014,Hockett2015a} for further details).
In the following section we explore this coherent control methodology
in terms of the intra-pulse dynamics, and further considerations for
applications to coherent metrology are discussed in the final section.

\section{Photoelectron interferograms from shaped pulses}

\begin{figure}
\includegraphics[scale=0.7]{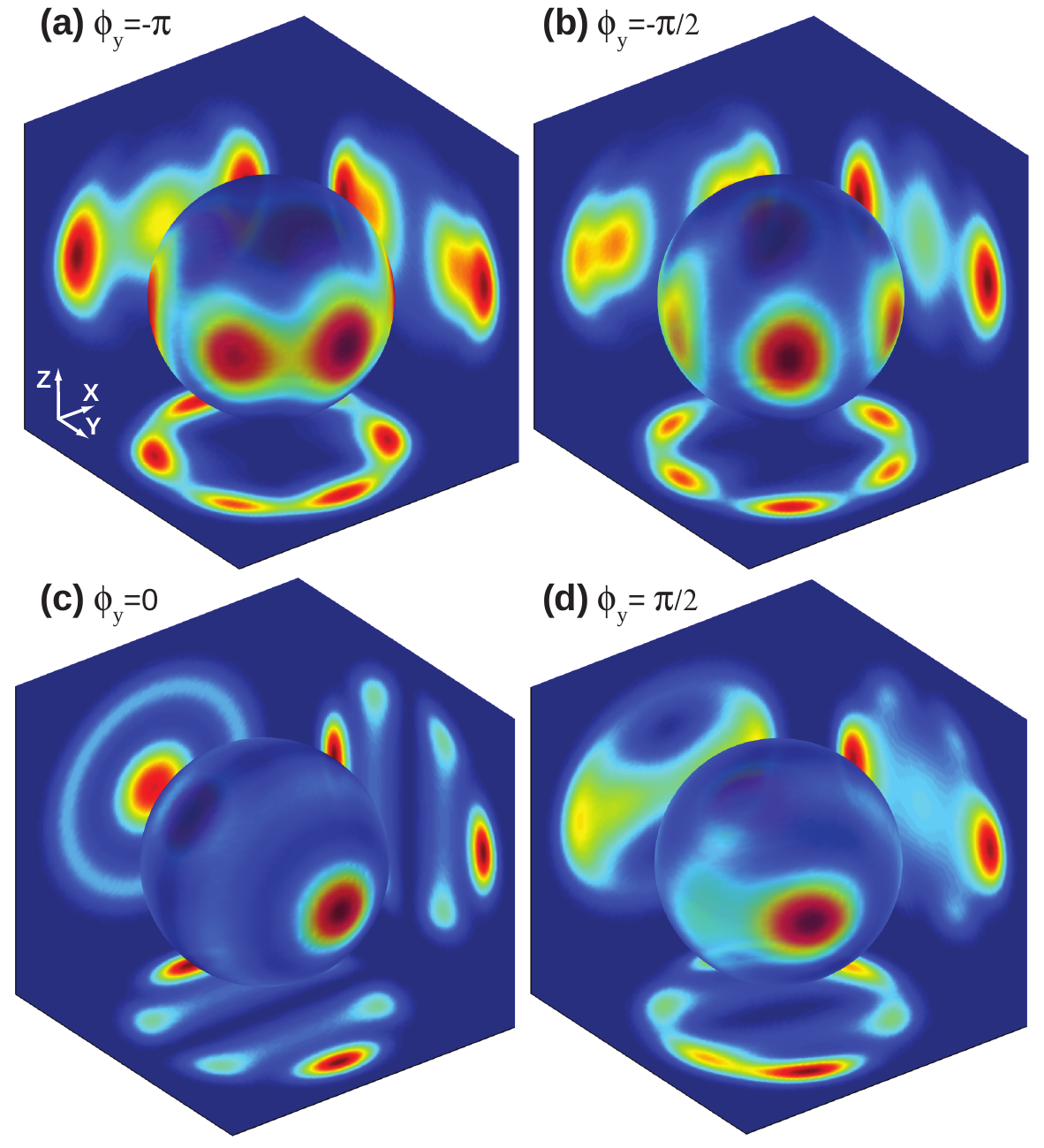}

\caption{Photoelectron interferograms from polarization-shaped laser pulses.
The angular interferograms $I(\theta,\phi;\, k)$ are shown on an
iso-sphere corresponding to a single $k$, and 2D projection planes
are shown assuming a Gaussian radial distribution $G(k)$ (see ref.
\cite{Hockett2015a}). The corresponding shaped laser pulses, $E(x,y,t)$,
parametrized by a spectral phase $\phi_{y}$, are shown in figure
\ref{fig:Polarization-shaped-pulses}. \label{fig:Photoelectron-interferograms-4up}}

\end{figure}

\begin{figure*}
\includegraphics[scale=0.95]{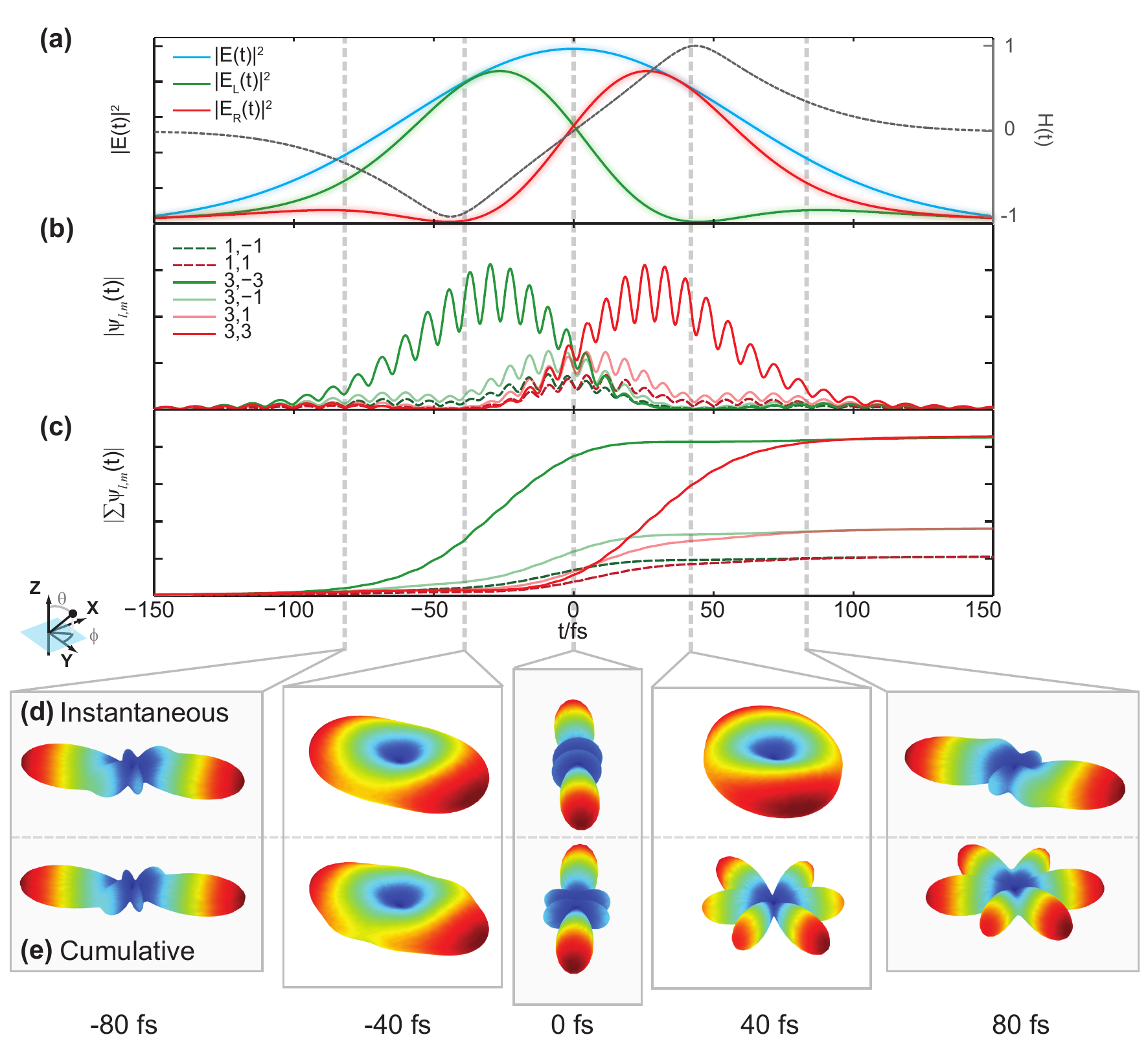}

\caption{Intrapulse continuum dynamics. (a) Shaped laser pulse for a spectral
phase step $\phi_{y}=-\pi$, expressed in the $L/R$ basis. This is
the same pulse as shown in figure \ref{fig:Polarization-shaped-pulses}(a).
(b) Instantaneous contribution to the continuum wavefunction $\Psi$,
for each partial-wave channel $\psi_{l,m}$. (c) Cumulative continuum
wavefunction. (d) Instantaneous and (e) cumulative photoelectron interferograms,
$\Psi^{*}\Psi$, plotted in polar form. The final cumulative result
corresponds to the interferogram shown in figure \ref{fig:Photoelectron-interferograms-4up}(a).
\label{fig:Intrapulse-continuum-dynamics}}

\end{figure*}

To illustrate the control of continuum wavepackets with polarization-shaped
pulses, figure \ref{fig:Photoelectron-interferograms-4up} shows examples
for four different pulse shapes. Here the pulse shapes are parametrized
as defined in eqn. \ref{eq:phi_y}, by the spectral phase $\phi_{y}$
which is applied to the red half of the pulse spectrum for the $E_{y}(t)$
component only. For $\phi_{y}=0$ the pulse is unchanged from its
initial linear polarization state, and a cylindrically symmetric distribution
is observed (fig. \ref{fig:Photoelectron-interferograms-4up}(c)),
while for all other cases the resultant continuum wavefunction is
more complicated, and reflects both the laser-driven bound-state dynamics
and ionization dynamics, integrated coherently over the pulse duration. 

To illustrate these dynamics, inherent to the control process, figure
\ref{fig:Intrapulse-continuum-dynamics} shows the temporal behaviour
of the continuum wavefunction. In this example, $\phi_{y}=-\pi$;
the resulting pulse shape was already shown in terms of the spatio-temporal
function $E(x,y,t)$ in fig. \ref{fig:Polarization-shaped-pulses}(a),
and is shown in terms of the $E_{L/R}(t)$ components in fig. \ref{fig:Intrapulse-continuum-dynamics}(a).
This pulse has a double-peaked structure in terms of the $E_{L/R}$
components of the field, hence a pulse which is close to pure circular
polarization states in temporal regions where one component dominates,
and passes through varying degrees of ellipticity as the $E_{L}:E_{R}$
ratio changes. This is shown most directly by the helicity $H(t)$
(eqn. \ref{eq:helicity}), plotted in panel (a) (see also fig. \ref{fig:Polarization-shaped-pulses}(a)).
Figure \ref{fig:Intrapulse-continuum-dynamics}(b) shows the amplitudes
of the instantaneous contributions to the continuum wavefunction,
$\psi_{l,m}(t)$. These amplitudes follow closely the $E_{L}(t)$
and $E_{R}(t)$ field components, since these components drive the
ionization towards $m<0$ and $m>0$ states respectively (see refs.
\cite{Hockett2014,Hockett2015a} for further details). At any given
instant $t$, the photoelectron interferogram is given by the coherent
square of the continuum wavefunction, hence will depend on the amplitudes
and phases of the $\psi_{l,m}(t)$ components, and a few examples
are shown in figure \ref{fig:Intrapulse-continuum-dynamics}(d). Similarly,
figure \ref{fig:Intrapulse-continuum-dynamics}(c) and (e) show the
cumulative continuum wavefunction, i.e. the coherent temporal sum
over $\psi_{l,m}(t)$ up to time $t$, and resultant photoelectron
interferograms. These plots therefore indicate the coherent evolution
of the final angular interferogram; as continuum population builds-up
over the laser pulse, the instantaneous and cumulative interferograms
diverge. In essence, these results highlight the sensitivity of the
final state wavefunction, and resultant interferogram, to the exact
shape of the pulse, and the inherent polarization-multiplexing.

In this example, the double-peaked nature of the pulse in the $E_{L/R}$
basis is particularly apt for consideration of the coherent build-up
of the continuum wavefunction. The instantaneous contribution to the
continuum for pure left and pure right circularly polarized light
are identical, aside from the sign of the $m$ terms prepared, which
only affects the handedness of the phase of $\psi_{l,m}$ with respect
to the angle $\phi$. Therefore, in ionization with a pure circular
polarization state, only final states with $l=3$ and $m=+3$ or $m=-3$
would be populated, and a $\phi$-invariant interferogram would result.
However, if both states are populated and interfere, the $\phi$-invariance
is broken due to the opposite phase of the states, which is given
by $e^{-im\phi}$. This is the essence of polarization-multiplexing,
in which continuum contributions correlated with ionization from different
polarization states can interfere. With a polarization-shaped pulse,
this interference occurs temporally, with the coherent build-up of
the continuum wavefunction over the pulse. In the case illustrated
here, the complexity is increased somewhat and there is significant
population of $m=\pm1$ (for both $l=1$ and $l=3$) states over parts
of the pulse, resulting in an observable with rich angular structure
(handedness and/or multiple lobes) for many of the instantaneous contributions,
as well as in the cumulative interferogram; despite this complexity
the underlying mechanism of temporal polarization-multiplexing is
conceptually identical to the simpler case described above. This process
is also functionally identical to the time-domain interferences observed
in the photoelectron energy spectrum following ionization by a double-pulse,
as discussed in ref. \cite{Wollenhaupt2002}. In that case the polarization
of both pulses was linear, but the addition of a temporal phase to
the electron wavepackets, via the use of two time-separated pulses,
created additional interferences which could be observed in the (time-integrated)
photoelectron energy spectrum. In all cases, the final observable
maintains coherence over the photoelectron wavepacket(s), allowing
for coherent control over this observable via the applied electric
field(s). Conceptually, all of these cases with doublet-structured
pulses are time-domain analogues of Young's double slit, in which
phase control is applied in the temporal rather than spatial dimension
\cite{Wollenhaupt2013,Wollenhaupt2014}, although the simple analogy
belies the rapid increase in complexity (beyond simple doublet structures)
which can be readily achieved in the time-domain.

\section{Coherent control for metrology}

\begin{figure}
\includegraphics[scale=0.85]{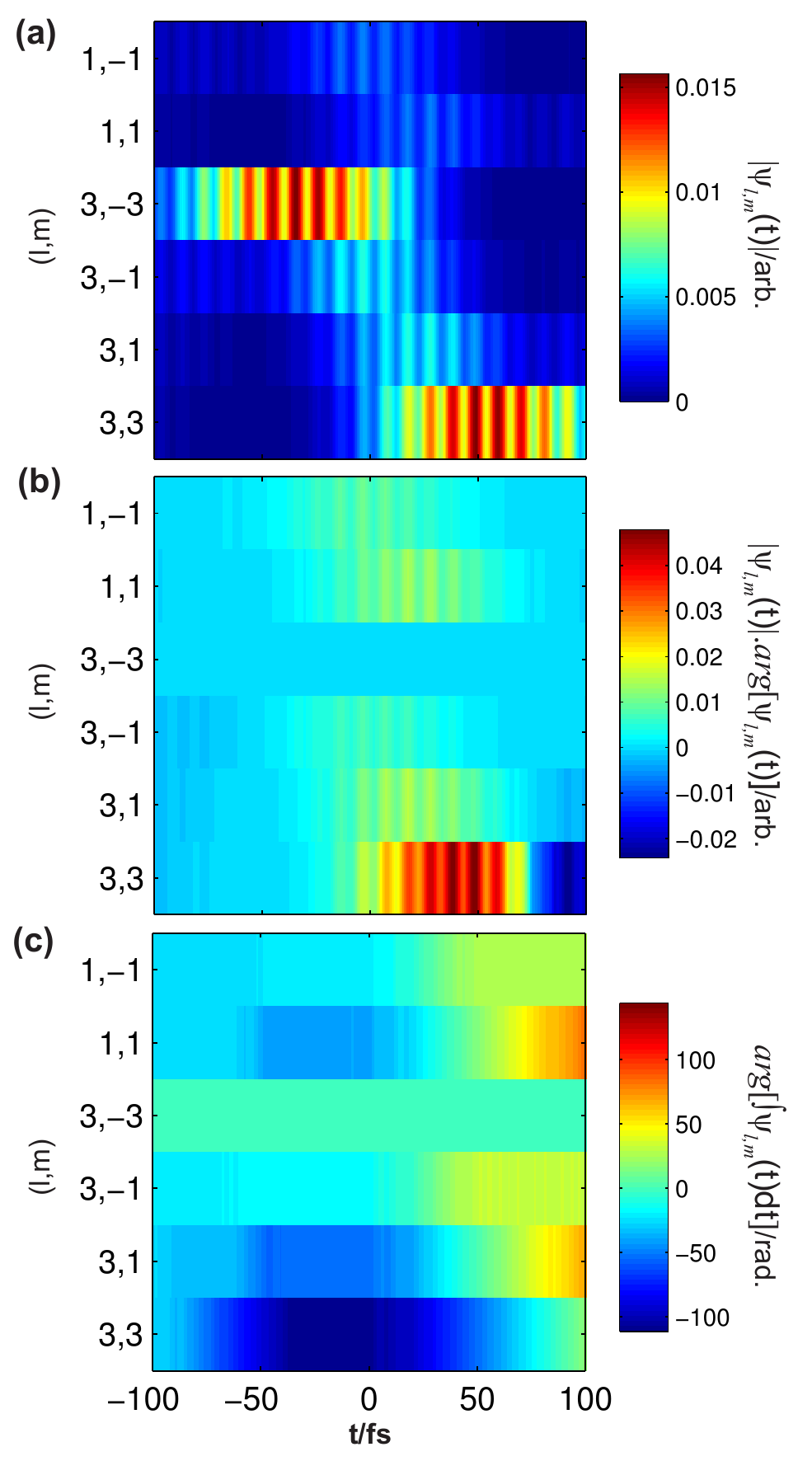}

\caption{Interferences in the continuum wavefunction. (a) magnitude of the
continuum wavefunction components $\psi_{l,m}(t)$, shown for all
contributing $(l,\, m)$ components; (b) instantaneous phase contributions,
$arg[\psi_{l,m}(t)]$, relative to the reference phase $arg[\psi_{3,-3}(t)]$
and weighted by the magnitude of the channel; (c) cumulative continuum
phase $arg[\int dt\psi_{l,m}(t)]$, relative to the reference phase
$arg[\int dt\psi_{3,-3}(t)]$. These results are for a shaped pulse
with $\phi_{y}=-\pi$, as shown in fig. \ref{fig:Polarization-shaped-pulses}(a).\label{fig:Interferences-continuum}}

\end{figure}

\begin{figure}
\includegraphics[scale=0.82]{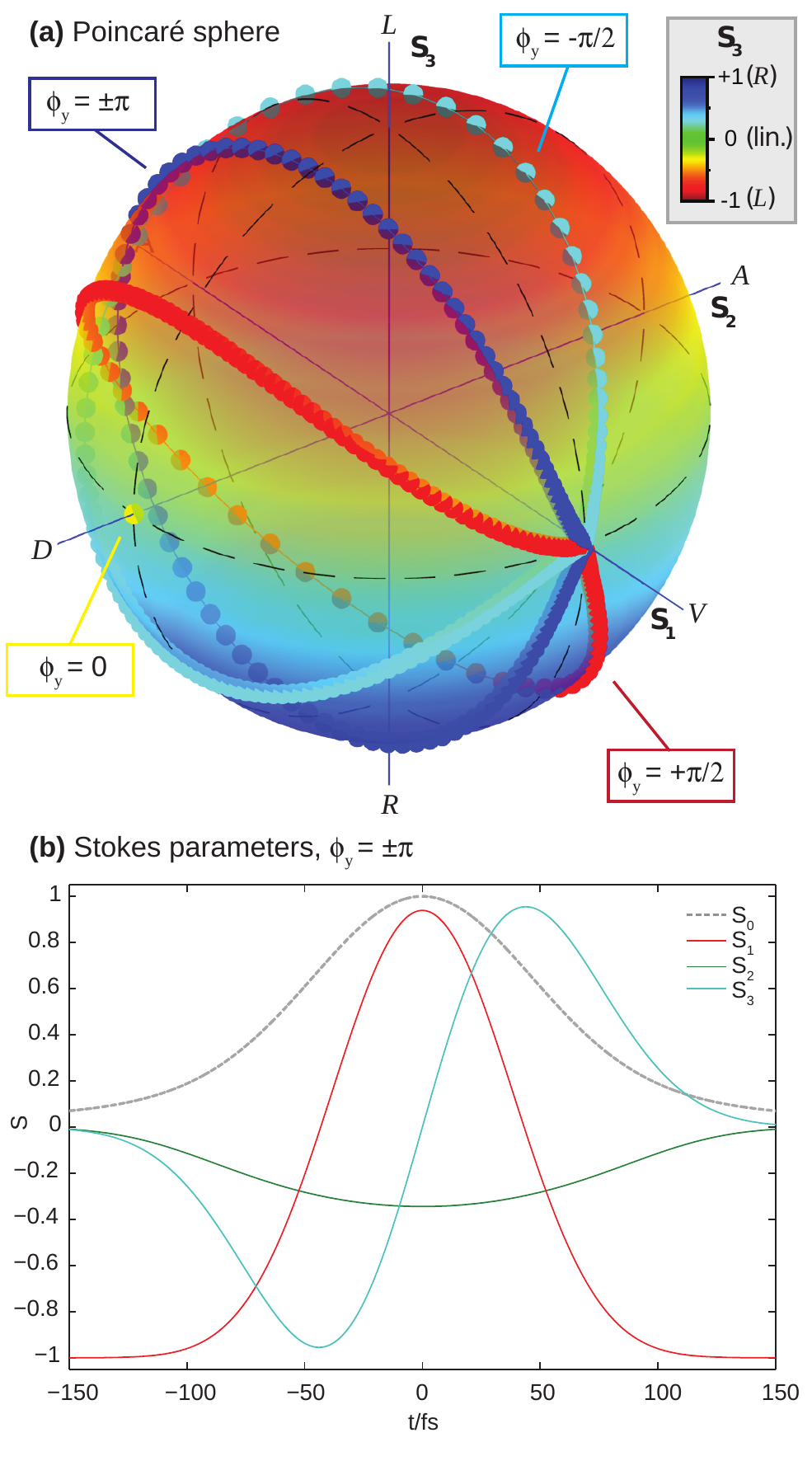}

\caption{Sampling paths in polarization space, defined by the time-dependent
Stokes parameters $(S_{1},\, S_{2},\, S_{3})$ and plotted on a Poincaré
sphere. (a) Poincaré sphere. Each path on the sphere corresponds to
a polarization-shaped pulse created by a spectral phase-step of the
type defined in eqn. \ref{eq:phi_y}, as shown in figure \ref{fig:Polarization-shaped-pulses},
with points plotted every 2.5~fs. The paths are parametrized by the
Stokes parameters, corresponding to the polarization basis states
$(H,\, V),\,(D,\, A)$ and $(L,\, R)$ respectively \cite{Guenther1990}.
The surface colour-map additionally indicates the $S_{3}$ value,
hence the pulse polarization in the $L/R$ basis. (b) Time-dependence
of the Stokes parameters $S_{n}(t)$ (here normalized by division
by $S_{0}(t)$) for the pulse defined by $\phi_{y}=\pm\pi$. \label{fig:Sampling-paths}}

\end{figure}

In the preceding section we focussed on the sensitivity of the final
continuum wavefunction to polarization-shaped pulses, in the case
where the details of the bound-continuum coupling are known. The opposite
also holds: this sensitivity can be used as a coherent quantum metrology
in order to discern details of the ionization dynamics. In fact, it
was this sensitivity that enabled the bound-continuum couplings to
be determined (see refs. \cite{Hockett2014,Hockett2015a}), although
polarization-shaped pulses were not exploited in that case. A natural
question which arises in this context is whether pulse-shapes can
be tailor-made for a particular measurement, and what benefit this
carries over a set of serial measurements with no polarization-multiplexing.
Although this possibility has been discussed in previous work in general
terms, the details have yet to be explored.

To approach this question, one must first consider the nature of the
metrology, and the information content required from an observable.
In this case we are concerned with ``complete'' photoionization
experiments, in which the bound-continuum couplings - the magnitudes
and phases of the ionization matrix elements - are determined \cite{Reid2003}.
Clearly, this information is available from the photoelectron interferograms,
since they are the coherent sum over all contributing continuum channels
$\psi_{l,m}$, hence sensitive to both the amplitudes and phases of
these channels. However, for any single polarization state, the interferences
present will be limited by the bound-continuum couplings inherent
to the light-matter system, so a single measurement will not usually
contain enough information to uniquely determine the full set of $\psi_{l,m}$
present. Therefore, in a serial measurement scheme, a set of interferograms
are obtained by varying the laser polarization (or via other means
of altering only the geometric aspects of the problem, see refs. \cite{Reid2003,Hockett2014,Hockett2015a}
and references therein for further discussion) are obtained, and analysed
globally to obtain the underlying properties. A polarization-multiplexed
measurement therefore represents a powerful alternative, since all
possible ionization channels may be accessed coherently, as a function
of time, over the pulse duration. This allows for the possibility
of rapid photoelectron metrology in general, and also for high-sensitivity
measurements by the choice of pulse-shapes which target certain channels
or interferences.  With respect to coherent control of chemical reactions,
this analysis underscores the findings that complex shaped laser pulses
can be more selective as compared to simple pulse shapes \cite{Assion1998,Brixner2001a},
and is confluent with suggestions to use specifically shaped light
fields for a spectroscopy ansatz going beyond the typical pulse sequence
spectroscopy with Fourier limited pulses \cite{Wohlleben2005,Wollenhaupt2011}. 

To highlight the potential of this coherent control scheme for metrology,
figure \ref{fig:Interferences-continuum} illustrates the relative
amplitude and phase contributions, as a function of time, for the
same $\phi_{y}=-\pi$ case discussed in the previous section. In these
results, the phases are referenced to the $\psi_{3,3}(t)$ channel,
and shown for both instantaneous contributions to the continuum wavepacket
$\psi_{l,m}(t)$, and cumulative continuum components $\int dt\psi_{l,m}(t)$.
In the instantaneous case, figure \ref{fig:Interferences-continuum}(b),
the results are weighted by the amplitudes of the channels to better
visualize the interferences present at any given $t$, while the cumulative
phase is shown without such weighting. This figure presents a more
comprehensive, but less intuitive, time-domain picture of the creation
of the continuum wavepacket than the discrete angular-interferograms
of figure \ref{fig:Intrapulse-continuum-dynamics}(d) and (e), since
it indicates the contributing interferences for all time-steps. As
discussed generally above, and in the previous section, these results
indicate how the various continuum channels build up coherently over
the laser pulse. Although the specific details are complicated, it
is clear that the instantaneous contributions sample many different
interferences - different relative magnitudes and phases between the
set of contributing $(l,\, m)$ channels - which coherently add over
the temporal coordinate to produce the final observable.

More explicitly, these results can be considered in terms of the information
content of the observable. In the case of measurements based on pulse
polarization, all possible measurements exist in the space of all
possible pulse polarizations, $\{H\}$. A time-invariant helicity,
denoted $\bar{H}(\bar{\phi_{y}})$, corresponds to a single point
in this measurement space, parametrized by the spectrally-invariant
phase $\bar{\phi_{y}}$, while a time-dependent helicity $H(t;\,\phi_{y}(\Omega))$,
parametrized by a spectral phase function $\phi_{y}(\Omega)$, samples
a sub-set of points within this space. Figure \ref{fig:Sampling-paths}
illustrates this sampling of the pulse polarization space, expressed
in terms of time-dependent Stokes parameters $S_{n}(t)$ (normalized
by $S_{0}(t)$) and plotted on a Poincaré sphere. Here the paths on
the sphere show the time-dependent sampling by a polarization-shaped
pulse created by a spectral phase-step of the type defined in eqn.
\ref{eq:phi_y}, for the pulses shown in figure \ref{fig:Polarization-shaped-pulses}.
For the case of $\phi_{y}=0$ there is no time-dependence to the pulse
polarization, hence only a single point in the polarization space
is sampled, while for all other $\phi_{y}$ a range of polarization
states are sampled.

To quantify the information content of a measurement, we can therefore
consider the size of the measurement space sampled by a single measurement,
or set of measurements. The information content of a single measurement
with a single polarization state is given as $M(\bar{H}(\bar{\phi_{y}}))$,
and a set of serial measurements therefore has an information content
of $M_{s}=\sum_{\bar{\phi_{y}}}|M(\bar{H}(\bar{\phi_{y}}))|$, where
the modulus is used to emphasize the fact that each measurement is
incoherent with respect to all others. Similarly, a single polarization-multiplexed
measurement has an information content given by $M_{p}=|\int dt\, M(H(t;\,\phi_{y}(\Omega)))|$,
and the modulus is used to emphasize the coherent nature of the parallel
case. A set of parallel measurements has information content $(M_{p})_{s}=\sum_{\phi_{y}(\Omega)}M_{p}=\sum_{\phi_{y}(\Omega)}|\int dt\, M(H(t;\,\phi_{y}(\Omega)))|$,
where each independent measurement corresponds to a laser pulse defined
by a spectral phase function $\phi_{y}(\Omega)$, and is incoherent
with respect to all other pulses. 

For an incoherent process, the information content of $M_{s}$ and
$M_{p}$ are equivalent for the case where $H(t;\,\phi_{y}(\Omega))=\sum_{\bar{\phi_{y}}}\bar{H}(\bar{\phi_{y}})$,
i.e. the time-dependent pulse samples the same set of points in the
measurement space as sampled by the set of measurements $M_{s}$.
Conceptually, this would correspond to the case where the instantaneous
photoelectron interferograms (figure \ref{fig:Intrapulse-continuum-dynamics}(d))
are summed incoherently, hence a set of serial measurements of these
interferograms would be identical to the result obtained with the
shaped-pulse. In this case, we can define an advantage in terms of
the scaling of the measurement time with the size of the polarization
sub-space: there is an advantage to a parallel measurement of $(N(\bar{\phi_{y}})-1)T$,
where $N(\bar{\phi_{y}})$ is the number of measurements sampled,
and $T$ the time required per measurement, assumed to be the same
regardless of the complexity of the pulse structure. 

However, for a coherent process, the information content of the final
result for the shaped-pulse case benefits from the fact that the coherent
addition over the measurement sub-space is massively parallel, with
many additional interferences accessed - this is truly a multiplexing
advantage, conceptually equivalent to the massive parallelization
inherent to quantum computing. Figure \ref{fig:Interferences-continuum}
visualizes this multiplexed process, since each temporal slice represents
a specific point within the $H$-space (i.e. one point on the sampling
path shown in figure ). Figure \ref{fig:Interferences-continuum}(b),
which shows the time-dependent contributions to the interferogram
for each continuum channel, relative to the reference channel, indicates
the specifics of each point in the sampled $H$-space for any given
channel (similarly, figure \ref{fig:Intrapulse-continuum-dynamics}(d)
shows the resulting angular-interferograms for a few of these points).
Here, both positive and negative contributions are observed relative
to the reference, indicating the sampling of points in $H$-space
with different relative phases. Figure \ref{fig:Interferences-continuum}(c)
shows the cumulative effect for the coherent temporal addition of
these $H$-space points on the relative phase of each channel, again
showing how the various channels can pass through regions in $H$-space
with constructive or destructive interferences as the continuum wavepacket
builds up, hence each point sampled can affect the final interferogram
distinctly.

In fact, in this particular case, many of these interferences could
not be accessed by serial measurement schemes since, by definition,
they cannot sample interferences between different points in $H$-space
(i.e. different points on the Poincaré sphere of figure \ref{fig:Sampling-paths}).
In order to sample such points, a serial measurement scheme based
on interfering the photoelectron wavepackets produced by two pulses
of different helicity would be required, such that the information
content would be defined by the sum over pairs of helicities: $M_{s*}=\sum_{\bar{\phi_{y}},\bar{\phi_{y}}'}|M(\bar{H}(\bar{\phi_{y}}))M(\bar{H}(\bar{\phi_{y}'}))|$.
In this case, the scaling law for measurement time reflects the square
of the sub-space sampled, $N(\bar{\phi_{y}})N(\bar{\phi_{y}}')T=N(\bar{\phi_{y}})^{2}T$.
However, since the parallel case maintains coherence over all measurement
points, a single measurement with a shaped-pulse which samples the
same sub-space, i.e. $H(t;\,\phi_{y}(\Omega))=\sum_{\bar{\phi_{y}}}\bar{H}(\bar{\phi_{y}})$
as defined above, has the same information content, and is obtained
in a measurement time $T$. Here the measurement time saving will
scale on $\mathcal{O}(N^{2})$, with value $(N(\bar{\phi_{y}})^{2}-1)T$.
Hence, coherent control in the time-domain represents a powerful tool
for metrology, with a high information content measurement obtainable
in a short measurement time, as compared to serial measurement schemes.

\section{Conclusions}

Photoelectron angular interferograms are a high information-content
observable. Coherent control over continuum photoelectron wavepackets
via the use of polarization-shaped laser pulses translates to a high
degree of control over this observable, while a deep understanding
of this observable provides a means to a coherent quantum metrology
for photoionization processes, including dynamics, with an inherent
multiplexing advantage.

Here we have explored these points by analysing, in detail, the computational
results and specific details of a single polarization-shaped pulse,
and used this to highlight the general concepts. The information content
of such measurements was investigated, in contrast to measurements
with pulses of a single polarization state, and seen to provide a
significant multiplexing advantage in terms of the inherent information
content of the measurement, and the measurement time required to obtain
the same information content via a set of serial measurements.

\section{Acknowledgements}

We thank R. Lausten for helpful discussion. Financial support by the
State Initiative for the Development of Scientific and Economic Excellence
(LOEWE) in the LOEWE-Focus ELCH is gratefully acknowledged.

\bibliographystyle{unsrt}
\bibliography{baumert_paper4_240415}

\end{document}